\documentclass[5p,twocolumn]{elsarticle}

\usepackage{graphicx}
\usepackage{dcolumn}
\usepackage{pifont}
\usepackage{bm}
\usepackage{multirow}
\usepackage{amsmath}
\usepackage{float}
\usepackage{txfonts}
\usepackage[version=3]{mhchem}

\journal{Nano Energy}

\begin{document}
\title{Mixed eldfellite compounds \ce{Na(Fe_{1/2}M_{1/2})(SO4)2} (M = Mn, Co, Ni): A new family of high electrode potential cathodes for the sodium-ion battery}

\author[kimuniv-m]{Gum-Chol Ri}
\author[kimuniv-m]{Song-Hyok Choe}
\author[kimuniv-m]{Chol-Jun Yu\corref{cor}}
\ead{ryongnam14@yahoo.com}
\cortext[cor]{Corresponding author}

\address[kimuniv-m]{Department of Computational Materials Design, Faculty of Materials Science, Kim Il Sung University, \\ Ryongnam-Dong, Taesong District, Pyongyang, Democratic People's Republic of Korea}

\begin{abstract}
Natural abundance of sodium and its similar behavior to lithium triggered recent extensive studies of cost-effective sodium-ion batteries (SIBs) for large-scale energy storage systems. A challenge is to develop electrode materials with a high electrode potential, specific capacity and a good rate capability. In this work we propose mixed eldfellite compounds \ce{Na_x(Fe_{1/2}M_{1/2})(SO4)2} (M = Mn, Co, Ni) as a new family of high electrode potential cathodes of SIBs and present their material properties predicted by first-principles calculations. The structural optimizations show that these materials have significantly small volume expansion rates below 5\% upon Na insertion/desertion with negative Na binding energies. Through the electronic structure calculations, we find band insulating properties and hole (and/or electron) polaron hoping as a possible mechanism for the charge transfer. Especially we confirm the high electrode voltages over 4 V with reasonably high specific capacities. We also investigate the sodium ion mobility by estimating plausible diffusion pathways and calculating the corresponding activation barriers, demonstrating the reasonably fast migrations of sodium ions during the operation. Our calculation results indicate that these mixed eldfellite compounds can be suitable materials for high performance SIB cathodes.
\end{abstract}


\maketitle
\section{Introduction}
Nowadays lithium-ion batteries (LIBs) are ubiquitous in portable electronic devices and often used to power hybrid and electric vehicles due to their excellent storaging capability in a high energy density. However, with a requirement of large-scale energy storage systems in exploiting natural energy sources such as solar and wind power, which is driven by ever increasing demand for energy and serious concern for environment, it has become clear that LIBs can not satisfy the increasing demand for battery due to a limited geographic location of lithium resources and thus a rapid rise in the price. In this context, considerable scientific effort has focused on finding alternatives to LIBs during the past decades. Recently, sodium-ion batteries (SIBs) based on the low-cost and unlimited sodium resources, which work in the same principle to LIBs, have attracted great attention as more suitable rechargeable batteries for large-scale grid applications~\cite{Palomares,Slater,Yabuuchi14, Sawicki,Wang15}.

Like LIBs, electrodes (cathode and anode) and electrolyte are the main components of SIBs, where sodium ion is inserted into and extracted from the electrodes reversibly during the charge-discharge process. However, the ionic radius of sodium ion (1.02 \AA) is larger than that of lithium ion (0.76 \AA), which makes it difficult to find suitable electrodes for sodium intercalation. Concerning the formula weight, moreover, sodium (23 g mol$^{-1}$) is about three times heavier than lithium (7 g mol$^{-1}$), possibly leading to a lower energy density for sodium versus lithium cells. Therefore, an extensive seeking and/or intensive computational design for new intercalation host materials with a high electrode potential and high power density is essential to achieve success in realization of commercially viable SIBs. This is in particular a great challenge for the cathode, due to preference of sodium for 6-fold coordination like octahedral or prismatic sites in crystalline materials. In this respect, the Na intercalated cathode hosts developed so far can be divided into two big classes, layered oxides and polyanionic materials.

The Na-based layered transition metal (TM) oxides with a general formula \ce{Na_xMO2} (M = transition metal) can be classified into O$n$-type (octahedral site) and P$n$-type (prismatic) according to the Delmas's notation~\cite{Delmas}, in which $n$ is the number of TM layers in the unit cell consisted of the \ce{MO6} edge-sharing octahedral units forming \ce{(MO2)}$_n$ sheets~\cite{Hasa1,Hasa2}. Although initial investigations for single TM layered oxides resulted in poor specific capacity and rate capability, the intermixing of TMs (mostly Mn, Fe, Co, Ni) in the \ce{TMO2}-layers improved the electrochemical performance and structural stability~\cite{Hasa1,Hasa2,Buchholz,Longo,Yuan,Pang,Vassil,DKim,Zhao,XLi,Qiao,Wen}. For instance, Li {\it et al.}~\cite{XLi} reported a synthesized \ce{Na(Mn_{0.25}Fe_{0.25}Co_{0.25}Ni_{0.25})O2}, which has a capacity of 180$-$239 mAh g$^{-1}$ and an average discharge voltage of 3.21 V, being suitable for the use in SIBs. Here, it is worth noting that Mn, Ni and Fe are of high elemental abundance, low cost and non-toxicity versus toxicity of Co, while Mn-, Co-, and Ni-based materials have redox process of \ce{M^{3+}}/\ce{M^{4+}} versus \ce{Fe^{2+}}/\ce{Fe^{3+}} in Fe-based cathodes. On the other hand, polyanionic materials such as phosphates (\ce{NaMPO4}) and fluorophosphates (\ce{NaMPO4F}) have also been extensively studied during the past decades, but they exhibited relatively low electrode voltages due to the low electronegative polyanions~\cite{Zhang,Moreau,JKim,Hautier}. In recent years, iron-based sulfate polyanionic compounds such as \ce{NaFe(SO4)2}~\cite{yucj17_2,Singh,ReynaudPhd}, \ce{Na2Fe2(SO4)3}~\cite{Barpanda14}, \ce{Na2Fe(SO4)2}~\cite{Reynaud14}, \ce{Fe2(SO4)3}~\cite{Mason} and \ce{Na2Fe(SO4)2\cdot}2\ce{H2O}~\cite{Meng} have been developed, which show relatively high electrode voltages of e.g., 3.8 V in \ce{Na2Fe2(SO4)3}~\cite{Barpanda14} and reasonably high energy densities due to the high electronegativity of the sulfate ion \ce{SO4^2-}. However, the electrode potential (and the specific capacity) of SIBs using polyanionic compound as cathode is still unattainable to the typical values over 4 V of LIBs.

In this work we propose eldfellite-based cathodes of SIBs with high electrode voltages and thus high specific capacities, newly designed by applying first-principles methods. The mixing approach is applied to the Fe site in eldfellite \ce{NaFe(SO4)2} with nearby TMs at the same low in the periodic table, resulting in the materials with formula \ce{Na_x(Fe_{1/2}M_{1/2})(SO4)2} (M = Mn, Co, Ni). Here we regard that the relatively low capacity of $\sim$80 mAh g$^{-1}$ in eldfellite \ce{Na_xFe(SO4)2}~\cite{yucj17_2,Singh} is mainly due to that Fe typically has ionic states of 2+ and 3+, i.e., \ce{Fe^{2+}}/\ce{Fe^{3+}} redox couple, and thus Na should be inserted into \ce{NaFe(SO4)2} host to become \ce{Na_{1+x}Fe(SO4)2}. On the contrary, Mn (or Co, Ni) has ionic states of 3+ and 4+, and therefore, if some of Fe atoms would be replaced by Mn (or Co, Ni), \ce{Mn^{3+}}/\ce{Mn^{4+}} redox reaction could be realized with an extraction of Na from \ce{Na(Fe_{1/2}Mn_{1/2})(SO4)2} to become \ce{Na_x(Fe_{1/2}Mn_{1/2})(SO4)2} (0 $<$ x $<$ 1). Such mixing of TMs is thought to be general for enhancement of electrode performance, as in the case of the layered transition metal oxides. As Ceder {\it et al.}~\cite{Vassil} pointed out, moreover, the substitution of Fe in \ce{NaFeO2} with Co, Mn, Ni/Co or Ni/Mn provided high reversible capacities and great rate capabilities due to suppressing iron migration.

\section{Computational methods}
\begin{figure}[!t]
\centering
\includegraphics[clip=true,scale=0.38]{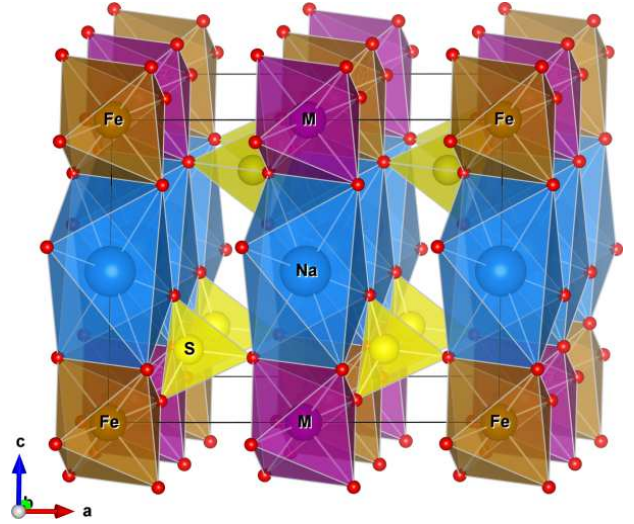}
\caption{\label{fig_str}Layered structure of $2\times2\times1$ supercell for eldfellite-based mixing compound \ce{Na(Fe_{1/2}M_{1/2})(SO4)2} (M = Mn, Co, Ni). Oxygen atoms are depicted by small red balls.}
\end{figure}

In respect of crystalline structure, monoclinic eldfellite \ce{NaFe(SO4)2} (space group; $C2/m$) is characterized by a layered structure, where the layer consists of of edge-sharing \ce{FeO6} and distorted \ce{NaO6} octahedra in the $ac$ plane or equivalently $bc$ plane~\cite{Zunic,ReynaudPhd}. These layers are connected by corner-sharing \ce{SO4} tetrahedra that leave interplanar space for two-dimensional \ce{Na+} guest ion diffusion. To allow the mixing of Fe site with Mn or Co or Ni, we construct supercells by doubling the primitive unit cell containing 1 formula unit (f.u.) in both [100] and [010] directions, {\it i.e.}, 2$\times$2$\times$1 supercells (4 f.u. = 48 atoms). Although different mixing ratios are thought to be possible, we only consider the compounds with the half of Fe atoms substituted with Mn or Co or Ni, whose chemical formula is \ce{Na(Fe_{1/2}M_{1/2})(SO4)2} (M = Mn, Co, Ni), as shown in Figure~\ref{fig_str}.

The pseudopotential plane-wave method within the density functional theory (DFT) framework as implemented in Quantum ESPRESSO package (version 5.3)~\cite{QE} is used to carry out the first-principles calculations. The ultrasoft pseudopotentials are used to describe the ion-electron interaction\footnote{We use the ultrasoft pseudopotentials Na.pbe-sp-van\_ak.UPF, Fe.pbe-sp-van\_ak.UPF, Mn.pbe-sp-van.UPF, Co.pbe-nd-rrkjus.UPF, Ni.pbe-n-rrkjus\_psl.0.1.UPF, S.pbe-van\_bm.UPF, and O.pbe-van\_ak.UPF from http://www.quantum-espresso.org.}, and the Perdew-Burke-Ernzerhof function within the generalized gradient approximation (GGA)~\cite{PBE} to treat the exchange-correlation (XC) interaction between the valence electrons. We adopt the DFT+$U$ method to reasonably deal with the localized $3d$ electrons of TMs with the on-site effective $U_\text{eff}$ ($U_\text{eff}=U-J$) parameters of 5.0, 4.5, 6.0 and 6.0 eV for Mn, Fe, Co and Ni, respectively~\cite{ZhouB1,Cococcioni}. The wave functions of valence electrons and electronic densities are expanded by the plane wave basis sets generated with the cut-off energies of 60 Ry and 480 Ry, respectively. Special $k$-points are set to be (2$\times$2$\times$4) for 2$\times$2$\times$1 supercells. These computational parameters guarantee a total energy accuracy of 5 meV per formula unit. The positions of all atoms and lattice parameters are fully relaxed until the atomic forces converge to 0.02 eV \AA$^{-1}$. Spin polarizations are considered in the structural optimization. As reported in our previous calculations~\cite{yucj17_2}, the ferromagnetic (FM) configuratiosn turn out to be slightly higher in total energy than the antiferromagnetic (AFM) configurations, and thus only the calculations for the AFM phases are used for an analysis of their electrochemical properties.

We calculate the activation energies for sodium ion migrations by using the climbing image nudged elastic band (NEB) method~\cite{NEB} as implemented in the neb code of the package. During the NEB run, the supercell lattice parameters are fixed at the optimized values, while all the atoms are allowed to relax. The seven image points are used and the convergence criteria for the force orthogonal to the path is set to be 0.05 eV \AA$^{-1}$. The $U_\text{eff}$ parameters are set to be zero and spin polarizations are not considered, since the effects on the activation energy is proven to be negligible.

\section{Results and discussion}
We first presented the lattice structures of monoclinic bulk crystals \ce{Na_x(Fe_{1/2}M_{1/2})(SO4)2} (M = Mn, Co, Ni) according to the sodium insertion ratio x, optimized by the GGA+$U$ method with spin-polarization, and calculated the formation and binding energies to discuss stabilities of the designed compounds. Based on the consideration that the Na atom can be extracted from \ce{Na(Fe_{1/2}M_{1/2})(SO4)2} as well as inserted into that, the sodium insertion ratios considered in this work are x = 0, 0.25, 0.5, 0.75, 1.0, 1.25, 1.5, and 1.75. For the cases of Na-inserted structures (x = 1.25, 1.5, 1.75), the analysis of bond valence sum difference ($\Delta$BVS) is carried out to estimate the possible positions of the inserted Na atoms~\cite{Brown_BVS,yucj17_2}. In Table~\ref{tab_lattice}, we show the lattice constants ($a=b, c$), relative volume expansion rate $r_\text{vol}=(V_x-V_{1.0})/V_{1.0}\times100$\%, formation and binding energies in the 2$\times$2$\times$1 supercells. The lattice angles are almost unchangeable according to both a kind of transition metal and sodium insertion ratio, as $\alpha=\beta=92^\circ, \gamma=66^\circ$. We note that in the case of eldfellite \ce{NaFe(SO4)2} the calculated lattice constants with the same computational parameters in our previous work~\cite{yucj17_2} were in good agreement with the experimental values~\cite{Zunic}.
\begin{table}[!b]
\caption{\label{tab_lattice}Calculated lattice constants, relative volume expansion rate $r_\text{vol}=(V_x-V_{1.0})/V_{1.0}\times100$\%, sodium binding energy $E_\ce{b}$, and formation energy $E_\ce{f}$ in 2$\times$2$\times$1 supercells for \ce{Na_x(Fe_{1/2}M_{1/2})(SO4)2}.}
\begin{tabular}{cccccc}
\hline
x & $a, b$ (\AA) & c (\AA) & $r_\text{vol}$ (\%) & $E_\ce{b}$ (eV) & $E_\ce{f}$ (eV)   \\
\hline
 & \multicolumn{5}{l}{\ce{Na_x(Fe_{1/2}Mn_{1/2})(SO4)2}} \\
0.00 & 9.53 & 7.06 & $-$1.01 & $-$     & 4.29    \\
0.25 & 9.52 & 7.20 & 0.82    & $-$5.59 & 3.18    \\
0.50 & 9.49 & 7.21 & 0.67    & $-$5.56 & 2.10    \\
0.75 & 9.50 & 7.20 & 0.56    & $-$5.51 & 1.04    \\
1.00 & 9.50 & 7.16 & 0.00    & $-$5.46 & $-$     \\
1.25 & 9.56 & 7.20 & 1.75    & $-$5.18 & $-$0.71 \\
1.50 & 9.64 & 7.18 & 3.03    & $-$4.96 & $-$1.39 \\
1.75 & 9.72 & 7.17 & 4.32    & $-$4.78 & $-$2.02 \\
\hline
 &\multicolumn{5}{l}{\ce{Na_x(Fe_{1/2}Co_{1/2})(SO4)2}} \\
0.00 & 9.51 & 7.07 & $-$0.85 & $-$     & 4.39    \\
0.25 & 9.50 & 7.19 & 0.81    & $-$5.69 & 3.26    \\
0.50 & 9.48 & 7.19 & 0.61    & $-$5.65 & 2.15    \\
0.75 & 9.49 & 7.18 & 0.57    & $-$5.61 & 1.07    \\
1.00 & 9.49 & 7.14 & 0.00    & $-$5.56 & $-$     \\
1.25 & 9.56 & 7.18 & 1.83    & $-$5.28 & $-$0.74 \\
1.50 & 9.64 & 7.17 & 3.15    & $-$5.07 & $-$1.45 \\
1.75 & 9.71 & 7.15 & 4.41    & $-$4.90 & $-$2.12 \\
\hline
 &\multicolumn{5}{l}{\ce{Na_x(Fe_{1/2}Ni_{1/2})(SO4)2}} \\
0.00 & 9.57 & 7.11 & $-$0.60 & $-$     & 4.43    \\
0.25 & 9.56 & 7.23 & 1.04    & $-$5.72 & 3.29    \\
0.50 & 9.56 & 7.22 & 0.60    & $-$5.69 & 2.17    \\
0.75 & 9.54 & 7.21 & 0.55    & $-$5.64 & 1.08    \\
1.00 & 9.54 & 7.18 & 0.00    & $-$5.61 & $-$     \\
1.25 & 9.60 & 7.21 & 1.60    & $-$5.33 & $-$0.76 \\
1.50 & 9.69 & 7.19 & 2.90    & $-$5.12 & $-$1.49 \\
1.75 & 9.76 & 7.18 & 4.19    & $-$4.96 & $-$2.18 \\
\hline
\end{tabular}
\end{table}

\begin{figure}[!th]
\centering
\includegraphics[clip=true,scale=0.5]{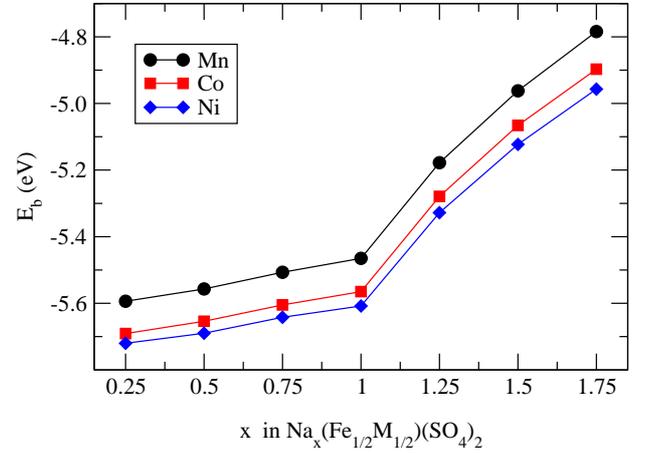}
\caption{\label{fig_eb}Sodium binding energy per Na atom in eldfellite-based mixing compounds \ce{Na_x(Fe_{1/2}M_{1/2})(SO4)2} (M = Mn, Co, Ni) as increasing the sodium concentration.}
\end{figure}

We calculate the binding energy of Na atom in the compounds to assess the interaction between Na atom and \ce{Na(Fe_{1/2}M_{1/2})(SO4)2} compounds. The binding energy per Na atom can be defined as follows,
\begin{equation}
E_\ce{b}=\frac{1}{x}\left[E_\ce{Na_x(Fe_{1/2}M_{1/2})(SO4)2} - E_\ce{(Fe_{1/2}M_{1/2})(SO4)2} - x E_\ce{Na_{gas}}\right] \label{eq_ebind}
\end{equation}
where $E_\ce{Na_x(Fe_{1/2}M_{1/2})(SO4)2}$ and $E_\ce{(Fe_{1/2}M_{1/2})(SO4)2}$ are the total energies of \ce{Na_x(Fe_{1/2}M_{1/2})(SO4)2} and \ce{(Fe_{1/2}M_{1/2})(SO4)2} compounds per f.u., and $E_\ce{Na_{gas}}$ is the energy of Na atom in gaseous state (isolated atom). All the negative binding energies, as shown in Table~\ref{tab_lattice}, indicate exothermic chemical interactions, and such relatively strong binding of Na atom implies high electrode voltage. The binding energy decreases in magnitude as increasing the Na concentration possibly due to a gradual enhancement of Na$-$Na interactions, while it increases going from Mn to Co and to Ni at each of Na concentration, as clearly shown in Figure~\ref{fig_eb}.

The relative stability of various eldfellite-based compounds upon Na extraction and insertion can be estimated by the formation energies for \ce{Na_x(Fe_{1/2}M_{1/2})(SO4)2} compounds with respect to \ce{Na(Fe_{1/2}M_{1/2})(SO4)2} compounds and body centered cubic (bcc) Na crystal using the following equation,
%
%\begin{multline}
%E_\ce{f}=E_\ce{Na_x(Fe_{1/2}M_{1/2})(SO4)2}- \frac{1}{2}(E_\ce{NaFe(SO4)2}+E_\ce{NaM(SO4)2}) \\ -(x-1)E_\ce{Na_{bcc}} \label{eq_eform1}
%\end{multline}
%where $E_\ce{Na_x(Fe_{1/2}M_{1/2})(SO4)2}$ is the total energy of \ce{Na_x(Fe_{1/2}M_{1/2})(SO4)2} compound per f.u., $E_\ce{NaFe(SO4)2}$ the energy of eldfellite \ce{NaFe(SO4)2} per f.u., $E_\ce{M_{bcc}}$ the energy per atom of M in the bcc crystal. As shown in Table~\ref{tab_lattice}, the calculated formation energies for all the compounds are negative, indicating a thermodynamic stability of the Na intercalated TM mixing compounds against the separation into most stable forms of eldfellite and bcc metals.
%
\begin{equation}
E_\ce{f}=E_\ce{Na_x(Fe_{1/2}M_{1/2})(SO4)2}-E_\ce{Na(Fe_{1/2}M_{1/2})(SO4)2}-(x-1)E_\ce{Na_{bcc}} \label{eq_eform}
\end{equation}
where $E_\ce{M_{bcc}}$ is the energy per atom of bcc crystalline sodium bulk. As shown in Table~\ref{tab_lattice}, the formation energies for 0 $<$ x $<$ 1 are calculated to be positive, indicating the endothermic formations of \ce{Na_x(Fe_{1/2}M_{1/2})(SO4)2} cpmpounds from \ce{Na(Fe_{1/2}M_{1/2})(SO4)2}, while those for 1 $<$ x $<$ 2 are negative. In fact, the former reaction is a charge process and the latter is a discharge process, provided that the \ce{Na(Fe_{1/2}M_{1/2})(SO4)2} compound is the reference state. We should note that the formation energies with respect to the \ce{(Fe_{1/2}M_{1/2})(SO4)2} compound are estimated to be negative, indicating that the Na intercalated compounds are thermodynamically stable. Moreover, $\Delta E=E_\ce{Na(Fe_{1/2}M_{1/2})(SO4)2}- \frac{1}{2}(E_\ce{NaFe(SO4)2}+E_\ce{NaM(SO4)2})$ according to general alloy theory are also calculated to be negative.

\begin{figure*}[!t]
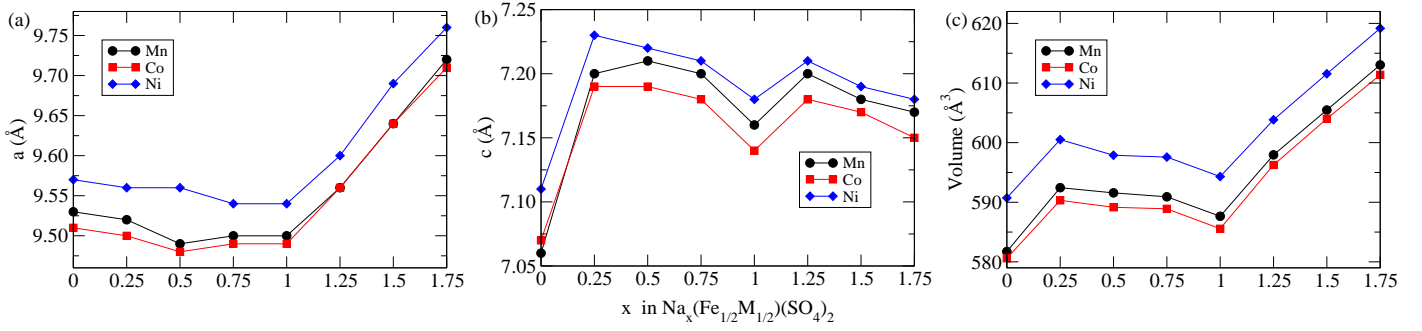

\centering
\includegraphics[clip=true,scale=0.35]{fig3a.eps}
\includegraphics[clip=true,scale=0.35]{fig3b.eps}
\includegraphics[clip=true,scale=0.35]{fig3c.eps}
\caption{\label{fig_lat}Variation of lattice constants $a$ (a) and $c$ (b), and cell volume (c) of $2\times2\times1$ supercells with Na concentration in \ce{Na_x(Fe_{1/2}M_{1/2})(SO4)2} (M = Mn, Co, Ni) compounds.}
\end{figure*}
It has been established that the electrode materials in general exhibit repetitive structural changes upon ion insertion and desertion during the battery operation, leading to capacity fade and a loss of electrochemical performance of the electrodes if the changes are serious or irreversible. To prevent this, the relative volume change rate has been suggested to be not over 5\% for polyanionic electrodes~\cite{Tripathi}. To check suitability of the suggested compounds for SIB cathodes, we have analyzed the variation of lattice parameters and volume when extracted or inserted Na atom, as shown in Table~\ref{tab_lattice} and Figure~\ref{fig_lat}. We find that the extraction of Na atom from \ce{Na(Fe_{1/2}M_{1/2})(SO4)2} (x $<$ 1) induces marginal expansion of lattice constants (maximally 0.04 \AA) and cell volume ($r_\text{vol}<1$\%), except the case of x = 0.0 at which $c$ and the volume are shrunk. This indicates that, upon the Na extraction, the attractive interaction between \ce{Fe^2+}/\ce{Fe^3+} and \ce{Na+} ions in the layer and the bridged sulfate \ce{SO4^2-} ions might be weakened, resulting in the enlargement of interlayer lattice constant $a$, while the repulsive interaction between \ce{FeO6} octahedra might be a little strengthened, leading to the slight increase of lattice constant $c$. When inserted Na atom into \ce{Na(Fe_{1/2}M_{1/2})(SO4)2} (x $>$ 1), the lattice constant $c$ changes in similar trend to the case of extraction, whereas the lattice constant $a$ increases distinctly with the maximum value of 0.22 \AA~at x = 1.75 and thus the volume expansion is also significant but still under 5\%. This can be interpreted that the inserted Na atom between layers enhances the interlayer repulsion due to additional \ce{Na+}$-$\ce{Fe^2+}/\ce{Fe^3+} and $-$\ce{Na+} repulsive interactions. We should emphasize that the volume expansion rates are quite small, especially for Na extraction, compared to over 20\% in the layered transition metal oxides, indicating a high reversible capacity. Among three kinds of mixing compounds, Ni mixing compounds have the largest volumes, while Co mixing the smallest volumes at all considered Na concentrations.

\begin{figure*}[!t]
\centering
\includegraphics[clip=true,scale=0.55]{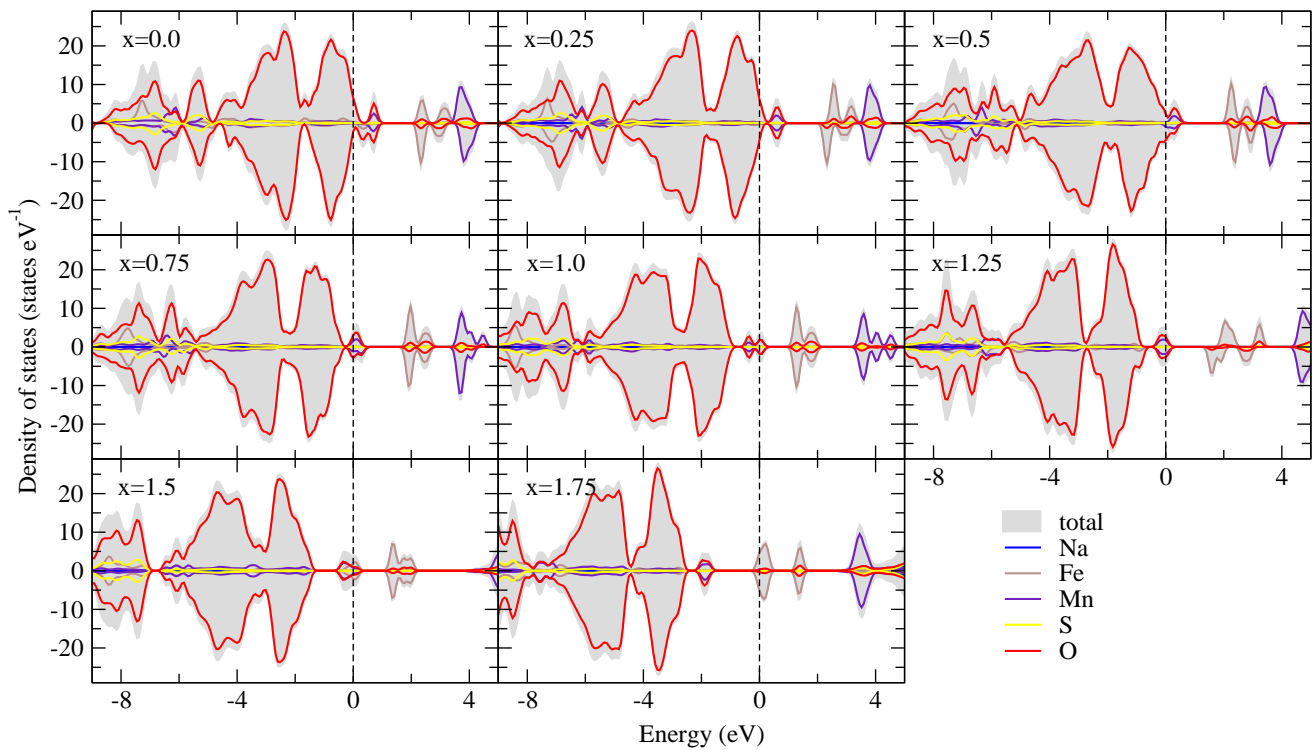}
\caption{\label{fig_dos}Atom-projected total density of states (DOS) for \ce{Na_x(Fe_{1/2}Mn_{1/2})(SO4)2} compounds as changing the Na concentration. The Fermi level is set to zero eV in each case.}
\end{figure*}

Next, we calculated the electronic structures of the compounds with a careful analysis to get an insightful understanding of electron transfer, which is of importance for SIB operation due to that although electrons travel through the external circuit, they merge with the \ce{Na+} ions in the cathode during the operation of the battery. Figure~\ref{fig_dos} shows the atom-projected density of states (DOS) for \ce{Na_x(Fe_{1/2}Mn_{1/2})(SO4)2} as increasing the Na concentration from X = 0.0 to 1.75. A distinct feature is that a small amount of states like impurity band state under x = 1.5, made from mostly O $p$ state, is found around the Fermi level, which is not the case for the pristine eldfellite \ce{NaFe(SO4)2}~\cite{yucj17_2}. This might be caused by solid solution effect between Fe and Mn (or Co, Ni). In spite of this feature, we can say that these Na intercalated compounds except the case of x = 1.75 are band insulators~\cite{MarianettiN}. As in other transition metal oxides, the valence bands are composed of oxygen 2$p$ electrons, whereas the conduction bands come from hybridized state between Fe 3$d$ and a little O 2$p$ states and Mn 3$d$ state at a short distance. Similar features were found for the cases of Co and Ni (see Figure S4). We note that the GGA+$U$ approach would not correct for oxygen atom contribution to conduction band energy level, and thus the metallic behavior might be overestimated if the conduction band edge has a large oxygen character. Upon the consideration that oxygen contribution to the conduction band edge in these compounds is not so much, we can retain our arguments to be acceptable. Then, we face with a problem how electrons transfer during electrode operation. In our previous work~\cite{yucj17_2}, we provided the charge transfer mechanism by electron polaron formation and migration upon the insertion of guest Na atom into the eldfellite \ce{NaFe(SO4)2}. Here we suggest that this mechanism is still valid in these mixed eldfellite, where the hole polaron instead of electron polaron is formed upon Na atom extraction from \ce{Na(Fe_{1/2}Mn_{1/2})(SO4)2} and migrates with a certain activation barrier. When sodium atom is extracted from \ce{Na(Fe_{1/2}Mn_{1/2})(SO4)2}, an extra hole is created in the compound since this leads to the removal of one electron (on the contrary, sodium insertion cause the creation of extra electron). This hole can be trapped by the Jahn-Teller type distortion of \ce{MO6} octahedra and thus form a hole polaron. According to our simple calculations, the negative self-trapping energy of about $-$0.56 eV was obtained, indicating that the hole polaron is stable, and the activation barrier for the polaron hoping was estimated to be $\sim$0.18 eV. Similar explanation for charge transfer has also been found in the Li-Ni-Co-Mn quaternary system that has been used as cathode of LIBs~\cite{Longo}. Similar arguments hold for M = Co, Ni as shown in Figure S1 and S2.

\begin{figure}[!b]
\centering
\includegraphics[clip=true,scale=0.23]{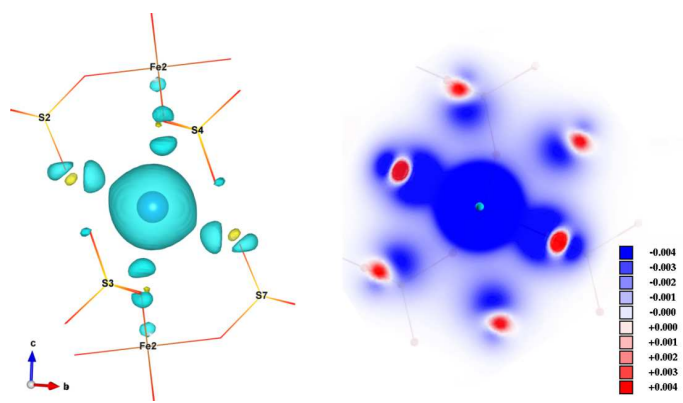}
\caption{\label{fig_dens}Charge density difference in \ce{Na_{0.25}(Fe_{1/2}Mn_{1/2})(SO4)2}, showing local region around Na atom. Light blue or blue color indicates electron loss, while yellow or red color indicates electron gain.}
\end{figure}

We analyze the net charge population of atoms obtained by applying Hirshfeld method~\cite{Hirshfeld} as implemented in SIESTA code~\cite{SIESTA}. Table~\ref{tab_hirsh} shows the average Hirshfeld net atomic populations.
\begin{table}[!t]
\caption{\label{tab_hirsh}Hirshfeld net atomic populations (average) for \ce{Na_x(Fe_{1/2}M_{1/2})(SO4)2} (M = Mn, Co, Ni) as increasing the Na concentration.}
\footnotesize
\begin{tabular}{cr@{\hspace{3pt}}r@{\hspace{3pt}}r@{\hspace{3pt}}r@{\hspace{3pt}}r@{\hspace{3pt}}r@{\hspace{3pt}}r@{\hspace{3pt}}r}
\hline
     & \multicolumn{8}{c}{Na concentration x} \\
\cline{2-9}
Atom & 0.0~ & 0.25~ & 0.5~~ & 0.75~ & 1.0~~ & 1.25~ & 1.5~~ & 1.75~ \\
\hline
Na &       & 0.089 & 0.097 & 0.097 & 0.095 & 0.106 & 0.109 & 0.111 \\
Fe & 0.229 & 0.228 & 0.231 & 0.230 & 0.233 & 0.236 & 0.241 & 0.246 \\
Mn & 0.244 & 0.245 & 0.245 & 0.247 & 0.247 & 0.251 & 0.257 & 0.263 \\
S  & 0.419 & 0.424 & 0.429 & 0.435 & 0.440 & 0.443 & 0.447 & 0.451 \\
O  &$-$0.134 &$-$0.138 &$-$0.143 &$-$0.147 &$-$0.152 &$-$0.158 &$-$0.163 &$-$0.169 \\
\hline
Co & 0.242 & 0.245 & 0.245 & 0.247 & 0.247 & 0.252 & 0.258 & 0.264 \\
Ni & 0.262 & 0.265 & 0.264 & 0.267 & 0.267 & 0.272 & 0.280 & 0.288 \\
\hline
\end{tabular}
\end{table}
Note that those for Co and Ni are from the corresponding mixed compounds, while those for other atoms only from TM = Mn. From the table, it is shown that only oxygen has negative net charge (electron excess), whereas all other atoms have positive net charge indicating a deficiency (donation) of electron. We emphasize that Na atom donates a large fraction of its valence electron indicating a strong ionic character of the bond between Na and the host. The amount of transferring electrons increases gradually as increasing the Na concentration. The nature of chemical bonding can also be described intuitively by plotting the charge density difference. As shown in Figure~\ref{fig_dens} for a typical distribution of the charge density difference in \ce{Na_{0.25}(Fe_{1/2}Mn_{1/2})(SO4)2}, Na atom and Fe atom lose their valence electrons, while O atoms accept the electrons.

\begin{figure}[!t]
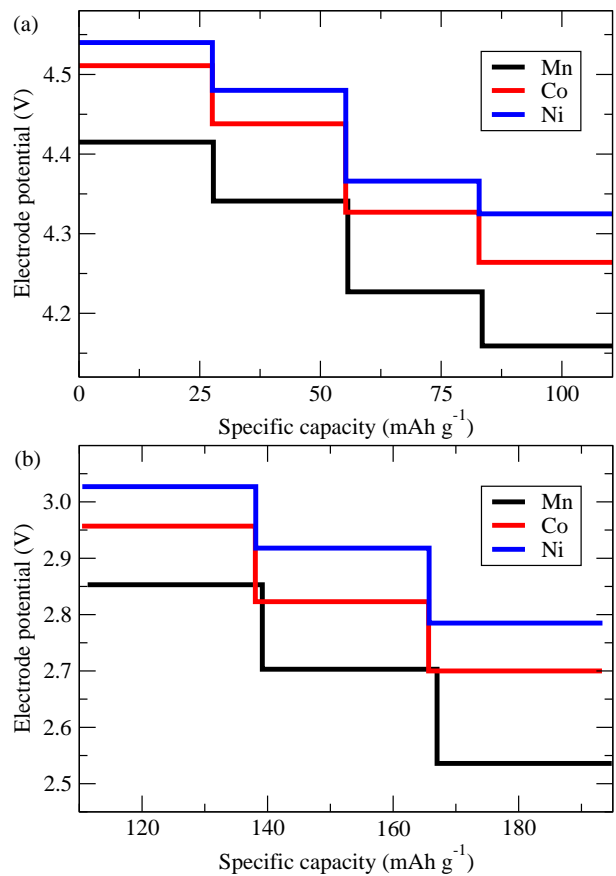

\centering
\includegraphics[clip=true,scale=0.5]{fig6a.eps}
\includegraphics[clip=true,scale=0.5]{fig6b.eps}
\caption{\label{fig_volt}Electrode potential as a function of specific capacity of \ce{Na_x(Fe_{1/2}M_{1/2})(SO4)2} (M = Mn, Co, Ni) compounds at (a) 0 $<$ x $<$ 1 for \ce{M^3+}/\ce{M^4+} discharge and (b) 1 $<$ x $<$ 2 for \ce{Fe^2+}/\ce{Fe^3+} discharge.}
\end{figure}

To shed light on the superior electrochemical properties of the designed compounds as powerful cathode materials for SIBs, we evaluate the electrode potential $V$ as a function of specific capacity. Here, the step discharge voltage between adjacent Na concentrations with respect to Na/\ce{Na+} counter electrode can be calculated as follows,
\begin{equation}
V = -\frac{E_{x_j}-E_{x_i}-(x_j-x_i)E_\ce{Na_{bcc}}}{e(x_j-x_i)}
\end{equation}
where $E_{x_j}$ is the total energy of the Na$_{x_j}$\ce{(Fe_{1/2}M_{1/2})(SO4)2} compound and $e$ the elementary charge. Figure~\ref{fig_volt} shows the reversible electrode potentials of the three kinds mixing materials as increasing the Na concentration from x = 0 to 1, which corresponds to the Mn (or Co, Ni) oxidation from \ce{Mn^3+} to \ce{Mn^4+}, and from x = 1 to 1.75 corresponding to the Fe oxidation from \ce{Fe^2+} to \ce{Fe^3+}. We find that the electrode potentials for the range of 0 $<$ x $<$ 1 vary in a range of 4.42$-$4.16 V with an average value of 4.23 V for Mn, 4.51$-$4.26 V with an average voltage of 4.33 V for Co, and 4.54$-$4.33 V with an average voltage of 4.37 V for Ni, respectively. These are remarkably high voltages compared with those of other polyanionic materials for SIB cathodes, and are comparable with those of typical cathodes of LIBs. These large magnitude of the concentration-dependent voltages can be ascribed to the strong binding of Na with the host compound. On the other hand, relatively low voltage profiles have been observed for the Na concentration 1 $<$ x $<$ 1.75 in Figure~\ref{fig_volt}(b); 2.85$-$2.54 V for Mn, 2.96$-$2.70 V for Co, and 3.03$-$2.79 V for Ni, respectively. These ranges of voltages are similar to those of eldfellite observed in the experiment~\cite{Singh} and computation~\cite{yucj17_2}, in which the guest Na atoms are additionally inserted into certain spatial positions of the host Na atom-containing eldfellite so that the \ce{Fe^2+}/\ce{Fe^3+} redox reaction is realized. The average voltage of Ni mixing compound is about 0.04 V higher than that of Co mixing compound, which is again 0.1 V higher than Mn mixing compound. These voltage difference according to mixing transition metal element can be attributed to stronger binding of Na with the host as shown in Table~\ref{tab_lattice}. From these results, it is verified that high electrode potential over 4 V can be realized for SIB cathode as well like LIB cathodes, provided that the redox couple of \ce{M^3+}/\ce{M^4+} would be found.

In Figure~\ref{fig_volt}, we can also see the maximum theoretical specific capacities of \ce{Na_x(Fe_{1/2}M_{1/2})(SO4)2} compounds to be about 110 mAh g$^{-1}$ if the charge/discharge process reverses at x = 1.0, and 194 mAh g$^{-1}$ if the process would extend to x = 1.75. Since it can be regarded that the extension of charge/discharge process might not be difficult as discussed later, these mixing materials between transition metals can posses also relative high specific capacity. In consequence, these mixing eldfellite materials can exhibit electrode voltage of 4.5$-$2.5 V with specific capacity up to 194 mAh g$^{-1}$, which is comparable with those of the binary$-$quaternary layered oxides~\cite{XLi}. When compared with other sulfate polyanionic material \ce{Na2Fe2(SO4)3}~\cite{Barpanda14}, the electrode voltage range is comparable such as 4.5$-$3 V but the specific capacity is higher if the extention to 1.75 is considered.

\begin{figure}[!b]
\centering
\includegraphics[clip=true,scale=0.34]{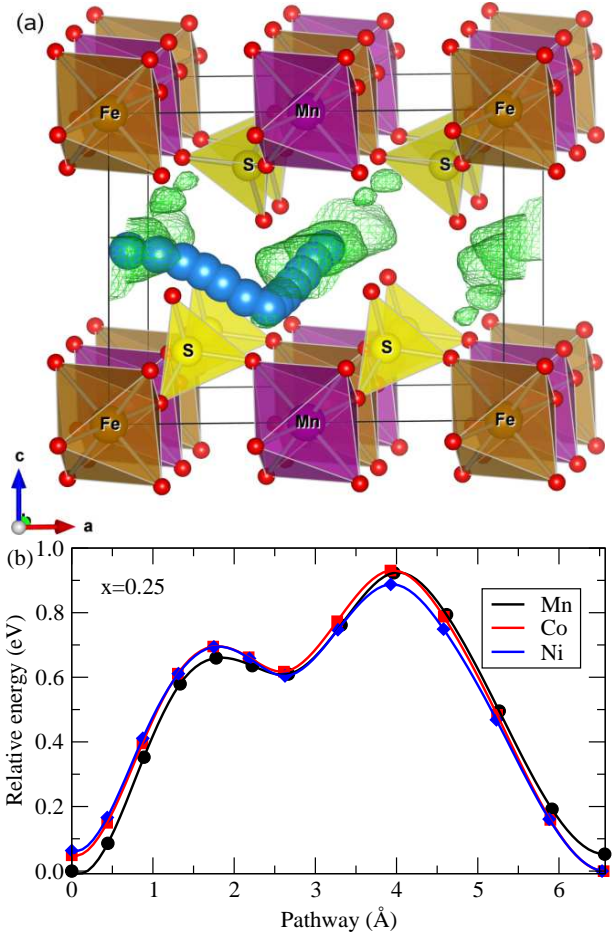} 
\includegraphics[clip=true,scale=0.5]{fig7b.eps}
\caption{\label{fig_nebene}(a) Isosurface plot of $\Delta$BVS at the value of 0.4 and intermediate positions of \ce{Na+} ion during the NEB simulation in \ce{Na_{0.25}(Fe_{1/2}Mn_{1/2})(SO4)2}, and (b) the corresponding energy profile of the three mixing eldfellite at x = 0.25. Na atoms are depicted by blue balls.}
\end{figure}
Finally we studied \ce{Na+} ion mobility, which determines one of the key factors for evaluating suitability of the electrode material such as the rate capability and cycling stability of the rechargeable SIBs. In order to estimate the mobility of \ce{Na+} ion, we have identified its migration pathway by applying $\Delta$BVS method, and calculated the activation barriers for \ce{Na+} ion diffusion along that pathway by using the NEB method. Considering that activation barriers depend on the concentration and the chemical environment, we first turned our attention to the case of low Na concentration, that is, x = 0.25. At this concentration, \ce{Na+} ion was regarded to migrate according to the vacancy-mediated mechanism, but it is still important to roughly estimate the migration pathway by calculating the $\Delta$BVS data and plotting in 3-dimension. Figure~\ref{fig_nebene}(a) shows the isosurface of the $\Delta$BVS in \ce{Na_{0.25}(Fe_{1/2}Mn_{1/2})(SO4)2} at the value of 0.4 and the intermediate positions of \ce{Na+} ion confirmed by the NEB calculation. As we have already found out in the pervious work~\cite{yucj17_2}, there are four identical interspace positions between layers identified by the $\Delta$BVS. The \ce{Na+} ion migrates through one of these positions to the vacancy at the same position of the adjacent cell in the $2\times2\times1$ supercell in this work. As shown in Figure~\ref{fig_nebene}, the local minimum in the energy is found at this position. The activation energies for Na migration from the start position to the local minimum position and subsequently from the local minimum position to the end position were calculated to be 0.66 eV and 0.27 eV in the Mn mixing eldfellite, respectively. The total activation energy of 0.87 eV is not really low, but it can be said to be reasonable compared with other phosphates, {\it e.g.}, marcite \ce{NaFePO4}~\cite{JKim} which exhibits $\sim$1.46 eV in crystalline phase and $\sim$0.73 eV in amorphous phase. This relatively high activation energy can be ascribed to influence of \ce{SO4} tetrahedra that are placed on the migration path. Reminding that the high electronegativity of the sulfate ion \ce{SO4^2-} can enhance the electrode potential, the sulfate ions have both positive and negative effects on the SIB performance. If the sulfate ions would shift from the migration path by either amorphization as in the case of \ce{NaFePO4} or other technique, \ce{Na+} ion diffusion could be expected to be easier. We then investigated \ce{Na+} ion diffusion as increasing the Na concentration. For x = 0.5 and x = 0.75, the migration paths are similar to the case of x = 0.25, while for x $>$ 1.0 we considered the effect of different configuration of inserted Na atom as we have conducted in the previous work~\cite{yucj17_2}. For x $>$ 1.0, in addition, we follow the mechanism of sequential move of host and guest Na atoms. The migration pathways and energy profiles for x = 1.5 and 1.75 are shown in Figure S3 and S4, respectively. Table~\ref{tab_neb} lists the activation energies as increasing the Na concentration.
\begin{table}[!t]
\caption{\label{tab_neb}Activation energy (eV) for Na atom migrations as increasing the Na concentration in \ce{Na_x(Fe_{1/2}M_{1/2})(SO4)2} (M = Mn, Co, Ni), calculated by the NEB method.}
\begin{tabular}{ccccccc}
\hline
     & \multicolumn{6}{c}{Na concentration x} \\
\cline{2-7}
TM & 0.25 & 0.5 & 0.75 & 1.25 & 1.5 & 1.75 \\
\hline
Mn & 0.87 & 0.95 & 0.94 & 0.86 & 0.55 & 0.36 \\
Co & 0.93 & 1.00 & 1.00 & 0.90 & 0.59 & 0.40 \\
Ni & 0.89 & 0.96 & 0.95 & 0.87 & 0.57 & 0.38 \\
\hline
\end{tabular}
\end{table}
The activation energy hits the highest at x = 0.5 and become lower drastically from x $>$ 1.5, arriving at the lowest value at x = 1.75 for all the mixing compounds. When comparing between Mn, Co and Ni mixing compounds, Mn compound has the lowest activation energy at all Na concentrations, while Co compound the highest value. This can be related with the net atomic charge of TM shown in Table~\ref{tab_hirsh}, where Co has the lowest charge and thus the weakest pushing effect on Na diffusion. When compared with the pure eldfellite, the activation energies are higher than those in the eldfellite, being associated with the smaller atomic charge of Fe than other TMs considered in this work.

\section{Conclusion}
In conclusion, we have designed a new family of cathode materials for SIBs with high electrode potentials, based on the eldfellite and mixing Fe with other transition metals including Mn, Co and Ni, resulting in the mixing compounds \ce{Na_x(Fe_{1/2}M_{1/2})(SO4)2} (M = Mn, Co, Ni). The DFT+$U$ method was applied through the work. We predicted their volume changes upon Na insertion/desertion, material stabilities by calculating formation energy and Na binding energy, electronic structures including the DOS and the net atomic charge, electrochemical properties such as electrode potential and specific capacity, and Na diffusion pathways and activation barriers. While demonstrating the very small volume change under 1\% relative volume expansion rate for 0 $\leq$ x $<$ 1 and 5\% for 1 $<$ x $\leq$ 1.75, the tailored mixing compounds have negative Na binding energies in range of $-$4.78 $\sim$ $-$5.73 eV varying with the Na concentration. From the analysis of the calculated DOS, these materials were turned out to be band insulators, and the hole polaron hoping was said to be the possible mechanism for charge transferring during battery operation. Most importantly, we have predicted remarkably high electrode voltages up to 4.37 V the average voltage for Ni mixing compound in the 0 $\leq$ x $<$ 1 Na concentration range, due to the \ce{Ni^3+}/\ce{Ni^4+} redox reaction. The combination of \ce{TM^3+}/\ce{TM^4+} and \ce{Fe^2+}/\ce{Fe^3+} redox couples, which can be easily realized, yielded the electrode voltage range of 4.54 $\sim$ 2.79 V with corresponding specific capacity up to 194 mAh g$^{-1}$. The NEB calculations produced the activation energies for \ce{Na+} ion diffusions of about 0.87 eV in Mn mixing compound at x = 0.25 with two-dimensional pathways. Although it is not possible at present to compare with experiment due to yet unexploit of these materials, we belive that the predicted materials properties should be confirmed by experiment in close future and this work will contribute to opening a new road for developing high performance cathode for commercially viable SIBs.

\section*{Acknowledgments}
This work is supported by the State Committee of Science and Technology, Democratic People's Republic of Korea, under the state project ``Design of Innovative Functional Materials for Energy and Environmental Application'' (No. 2016-20). The calculations have been carried out on the HP Blade System C7000 (HP BL460c) that is owned and managed by the Faculty of Materials Science, Kim Il Sung University.

\section*{Appendix A. Supporting information}
Supporting information related to this article can be found at URL.

\section*{\label{note}Notes}
The authors declare no competing financial interest.

\bibliographystyle{elsarticle-num-names}
\bibliography{Reference}

\end{document}